\documentclass[%
 reprint,
 amsmath,amssymb,
 aps,dvipdfmx
]{revtex4-1}
\usepackage{graphicx}
\usepackage{dcolumn}
\usepackage{bm}
\usepackage{color}

\renewcommand{\a}{\alpha}
\renewcommand{\b}{\beta}

\newcommand{\bea}{\begin{eqnarray}}
\newcommand{\eea}{\end{eqnarray}}
\newcommand{\f}[2]{\frac{#1}{#2}}
\newcommand{\eq}{&=&}
\newcommand{\nn}{\nonumber \\ }
\newcommand{\ve}{\varepsilon}

\newcommand{\area}{\int_{-\infty}^\infty }

\renewcommand{\l}{\lambda}
\newcommand{\p}{\partial}
\newcommand{\pp}[2]{\f{\p #1}{\p #2}}
\newcommand{\s}{\sigma}
\newcommand{\sref}[1]{Eq. (\ref{#1})}

\newcommand{\citeauthorname}[2]{{#1} {#2}}
\newcommand{\citebook}[4]{{#1} {\it #2} ({#3}, {#4}).}
\newcommand{\citepaper}[4]{{#1} {#2}, {#3} ({#4}).}
\allowdisplaybreaks[1]
\begin{document}

\preprint{APS/123-QED}

\title{Pythagorean theorem of Sharpe ratio
}

\author{Takashi Shinzato}
\email{Email address: takashi.shinzato@r.hit-u.ac.jp}
 \affiliation{
Mori Arinori Center for Higher Education and Global Mobility,
Hitotsubashi University, 
Tokyo 1868601, Japan}

\date{\today}

\begin{abstract}
In the present paper,
using {a replica analysis}, 
we {examine} the portfolio optimization problem handled in previous work and discuss 
the minimization of investment risk under constraints of budget and expected return for the case that 
the distribution of the hyperparameters of the mean and 
variance of the return rate of each asset are not limited to 
a specific probability family.
Findings derived using our proposed method are compared with those in previous 
{work}
to verify the effectiveness of our proposed method.
Further, we derive 
a Pythagorean theorem of the Sharpe ratio and 
macroscopic relations of opportunity loss. 
Using numerical experiments, 
the effectiveness of our proposed method is {demonstrated for a specific situation}. 
\begin{description}
\item[PACS number(s)]
{89.65.Gh}, {89.90.+n}, {02.50.-r}
\end{description}
\end{abstract}
\maketitle

\section{Introduction}
Nowadays, most financial activities 
interact with each other 
on a global scale and
our lives have been influenced, either 
directly or indirectly, by a number of financial crises.
The lessons of the financial crisis include the need to take personal effort to preserve our assets.
In this atmosphere and as reforms advance, 
the importance of making proper investments and managing risk 
has been recognized.
Generally speaking, 
investment means 
paying a cost in anticipation of future return and
often involves risk. Markowitz pointed out 
the importance of {investment} management and first laid out 
the portfolio optimization problem which is the 
framework for 
analyzing mathematically 
the optimal 
asset management strategy \cite{Markowitz1952,Markowitz1959}.
Several studies following this pioneering work
have been carried out \cite{Bodie,Luenberger,Prigent,Ang,Francis,Elton}.
Recently, there has been much such research 
that takes the viewpoint of 
complex systems
and actively applies analytical 
approaches refined in outside research fields, 
such as replica analysis, belief propagation, and 
random matrix theory, to the portfolio optimization problem
\cite{Ciliberti1,Ciliberti2,Caccioli,Pafka,Shinzato-2015-PLOS7,Shinzato-2015-PLOS8,Shinzato-2016-PRE12,Shinzato-2011-IEICE,Kondor-2016-with-no-short-selling,Shinzato-2017-JSTAT2,Shinzato-ve-fixed2016,Shinzato-2016-PRE11,VH}.

\begin{table*}[t]
{	\centering
{
\caption{Targets of previous studies.\label{tab1} In the context of statistical physics, the models are related to the Hamiltonian and
the constraints correspond to the priors.}
	\begin{tabular}{|p{4.5cm}|p{3.2cm}|p{3.5cm}|p{2.0cm}|p{3.5cm}|}\hline
		Researchers	&Model	&Constraints	&Optimizations	&Analysis approaches	\\ \hline
		Ciliberti, Ciliberti {\it et al.} \cite{Ciliberti1,Ciliberti2}	&absolute deviation model, expected shortfall model	&budget	&minimization	&replica analysis	\\ \hline
		Caccioli {\it et al.} \cite{Caccioli}	&expected shortfall mode, max loss model	&budget	&minimization	&replica analysis	\\ \hline
		Pafka {\it et al.} \cite{Pafka}	&mean-variance model	&budget	&minimization	&random matrix approach	\\ \hline
		Shinzato \cite{Shinzato-2015-PLOS7}	&mean-variance model	&budget	&minimization	&replica analysis	\\ \hline
		Shinzato {\it et al.} \cite{Shinzato-2015-PLOS8}	&any model	&budget	&minimization	&belief propagation	\\ \hline
		Shinzato \cite{Shinzato-2016-PRE12}	&mean-variance model	&budget	&minimization	&replica analysis, belief propagation	\\ \hline
		Shinzato, Kondor {\it et al.} \cite{Shinzato-2011-IEICE,Kondor-2016-with-no-short-selling}	&mean-variance model	&budget, short selling	&minimization	&replica analysis	\\ \hline
		Shinzato \cite{Shinzato-2017-JSTAT2}	&mean-variance model	&budget, investment concentration	&minimization	&replica analysis	\\ \hline
		Shinzato \cite{Shinzato-ve-fixed2016}	&mean-variance model	&budget, investment risk	&minimization and maximization	&replica analysis	\\ \hline
		Shinzato \cite{Shinzato-2016-PRE11}	&mean-variance model	&budget, expected return, investment risk	&minimization and maximization	&replica analysis	\\ \hline
		Varga-Haszonits {\it et al.} \cite{VH}	&a specific model	&budget, expected return	&minimization and maximization	&replica analysis	\\ \hline
	\end{tabular}
}
}
\end{table*}

{Among such research  (see Table \ref{tab1})}, Ciliberti {\it et al.} 
described 
a diversified investment system using the Boltzmann distribution 
to 
analyze the optimal portfolio 
for minimizing 
risk under a budget constraint. In particular, they
analyzed 
the minimal investment risk 
of the absolute deviation model 
and the expected shortfall model 
{using the ground state in the absolute zero temperature limit (that is, the optimal state of this optimization problem) \cite{Ciliberti1,Ciliberti2}.}
Moreover, Caccioli {\it et al.} 
examined 
the expected shortfall model 
with $L_2$ regularization 
and max loss model as a special case {of it}
by using replica analysis 
and identified the typical behavior of the optimal asset management strategy
\cite{Caccioli}.
Furthermore, Pafka {\it et al.}
discussed in detail 
the behavior of 
the investment risk
which is defined using 
the variance-covariance matrix
of the random weighted sums of
 each component 
of the lower triangular matrix 
which {can be extracted from} 
the true variance-covariance matrix with respect to the return rate of assets
by using Cholesky decomposition 
and 
the in-sample risk
by using the asymptotical spectrum of a random matrix which is 
generated by a given 
return rate \cite{Pafka}. Subsequently, 
Shinzato
analyzed 
one of the portfolio optimization problems, the mean--variance model, 
and 
showed that 
the minimal investment risk and its 
investment concentration satisfy 
the self-averaging property using the large deviation principle 
\cite{Shinzato-2015-PLOS7}.
Shinzato also
compared 
the minimal investment risk per asset 
derived using replica analysis 
with 
the minimal expected investment risk per asset 
derived using operations research 
from {a unified viewpoint of} stochastic optimization  
and pointed out 
that 
a portfolio which can minimize 
the expected investment risk
does not {necessarily}  minimize 
the investment risk. 
Furthermore, 
{Shinzato {\it et al.}}
developed a faster algorithm 
for finding 
the optimal portfolio 
which can minimize the risk function by using the belief propagation method, which is 
often used in probabilistic inference, 
and verified that 
the computation time of the algorithm is on the order of the square of the number of assets 
(whereas the standard algorithm requires on the order of the cube of the number of assets computation time).
Moreover, {they} also 
clarified 
that the Konno--Yamazaki conjecture which was previously 
confirmed in annealed disordered systems also 
holds true in quenched disordered systems \cite{Shinzato-2015-PLOS8}.
Additionally, 
Shinzato
{used}  the 
portfolio optimization problem 
of Ref.~\cite{Shinzato-2015-PLOS7}
and examined 
the portfolio which 
can minimize the 
investment risk 
under
a budget constraint for the case that 
the variances of the return rates of the assets are not unique
using replica analysis and belief propagation \cite{Shinzato-2016-PRE12}. 
Shinzato 
also investigated 
the minimization of investment 
risk under
constraints of budget and short selling
by using replica analysis 
and showed that this investment system involves a phase transition. Further, he 
developed 
a faster algorithm based on 
belief propagation for obtaining the optimal portfolio 
\cite{Shinzato-2011-IEICE}. 
In addition, Kondor {\it et al.}
analyzed 
the same portfolio optimization problem for the case that 
the variance of the return rate of each asset is 
distinct  
using replica analysis 
and reconfirmed 
that this disordered system involves a phase transition \cite{Kondor-2016-with-no-short-selling}.
Furthermore, 
Shinzato 
also {used}
the portfolio optimization problem handled in  
Ref.~\cite{Shinzato-2015-PLOS7} to 
examine 
the minimization 
of investment risk 
{under constraints of both budget and investment concentration }
by using replica analysis; in this context, 
he analyzed 
the optimization of 
investment concentration 
under constraints of budget and investment risk 
from {a unified viewpoint of stochastic optimization and duality} \cite{Shinzato-2017-JSTAT2,Shinzato-ve-fixed2016}.
In addition, Shinzato {used} the portfolio optimization problem handled in previous work 
\cite{Shinzato-2015-PLOS7} {to examine the minimization of investment risk under 
constraints of budget and }
expected return for the case that the variance of
the return rate is the same for all assets 
by using replica analysis \cite{Shinzato-2016-PRE11}. Moreover,
he also analyzed 
the maximization of expected returns
under constraints of budget and investment risk and 
pointed out the importance of 
duality for assessing these optimization problems.
Further, Varga-Haszonits {\it et al.}
examined 
the minimization of a particular risk function (the sample variance with respect to 
the deviation between the return and its sample average) under constraints of budget and expected return
by using replica analysis 
and carried out a stability analysis of the replica symmetric solution derived using replica analysis \cite{VH}.

As discussed above, several previous studies which further refined the model introduced in  
Ref.~\cite{Shinzato-2015-PLOS7}
have been reported 
\cite{Shinzato-2016-PRE12,Shinzato-2016-PRE11}. 
The findings of these studies are closely linked, 
which makes it possible to use them to solve 
an important problem. Namely, 
in Ref.~\cite{Shinzato-2016-PRE11}, 
the minimization of investment risk 
under constraints of budget and 
expected return for the case that the variance of the return rate of each asset is unique 
was discussed in detail, whereas 
in Ref.~\cite{Shinzato-2016-PRE12},
the minimization of investment risk
under a budget constraint 
for the case that the variances of the return rates of the assets are not unique 
was addressed.
As a natural {application of mathematical finance models}, 
we can also 
examine 
the minimization of investment risk 
under constraints of budget and 
expected return for the case that the variances of the return rates of the assets are not unique.
Moreover, 
in Ref.~\cite{Shinzato-2016-PRE11}, 
{hyperparameters of the means of the return rates of the assets 
are assumed to be independently and identically Gaussian distributed.
In this paper, }
following the above-described previous work,
we discuss 
the minimization of investment risk
under constraints of budget and expected return 
for the case that 
the 
distributions of 
the hyperparameters of the means and 
variances are not limited 
to 
a specific 
probability family 
and analyze 
the minimal investment risk per asset, 
investment concentration, and 
Sharpe ratio. 
Further, 
we derive a Pythagorean theorem of the Sharpe ratio and
macroscopic relations of opportunity loss 
along the lines of macro theory in mathematical finance (like thermodynamic relations).

This paper is organized as follows. 
In the next section, 
we formulate 
the 
minimization of investment risk 
under constraints of budget and expected return that is the focus of this study.
In section \ref{sec3},
we analyze 
the minimization of investment risk 
under these constraints 
by using replica analysis 
and derive the minimal investment risk, the investment concentration, and 
the Sharpe ratio.
In section \ref{sec4},
the results obtained using our proposed approach 
are {examined} in detail, 
and in section \ref{sec5}, 
we consider 
the validity of 
the proposed methodology 
using numerical simulations.
The final section gives 
our conclusions and lays out future work.

\section{Model setting}
In this study,
we consider 
a stable investment market which can handle 
$N$ assets without a restriction on short selling.
We assume that the return rates of assets $i(=1,2,\cdots,N)$, 
$\bar{x}_i$, 
are independently and identically distributed with 
mean $E[\bar{x}_i]=r_i$ and 
variance $V[\bar{x}_i]=v_i$. Moreover,
assuming $p$ investment periods, 
$\bar{x}_{i\mu}$ denotes 
the return rate of asset $i$ at period $\mu(=1,2,\cdots,p)$.
Furthermore, 
the portfolio of asset $i$ is $w_i\in{\bf R}$ and 
the portfolio of $N$ assets is described by $\vec{w}=(w_1,w_2,\cdots,w_N)^{\rm T}\in{\bf R}^N$, where 
the notation ${\rm T}$ means 
the transposition of a matrix or vector.
Similar to in previous work, 
since no restriction on short selling 
is imposed, 
the portfolio can take any real number.
In addition, the portfolio $\vec{w}$ is 
only under constraints of budget and expected return
\bea
\sum_{i=1}^Nw_i\eq N,\label{eq1}\\
\sum_{i=1}^Nw_ir_i\eq NR,\label{eq2}
\eea
respectively, 
where $R$ is the expected return coefficient.

Then, the investment risk of portfolio $\vec{w}$  
under these two constraints, ${\cal H}(\vec{w}|X)$, is defined as follows:
\bea
\label{eq3}
{\cal H}(\vec{w}|X)
\eq
\f{1}{2N}\sum_{\mu=1}^p
\left(\sum_{i=1}^Nw_i\bar{x}_{i\mu}-
\sum_{i=1}^Nw_ir_{i}
\right)^2\nn
\eq
\f{1}{2}\vec{w}^{\rm T}J\vec{w},
\eea
where the modified return rate $x_{i\mu}=\bar{x}_{i\mu}-r_i$ and
return rate matrix $X=\left\{\f{x_{i\mu}}{\sqrt{N}}\right\}\in{\bf R}^{N\times p}$ are used.
In addition, we will need matrix $J=\left\{J_{ij}\right\}=XX^{\rm T}\in{\bf R}^{N\times N}$, which has $i,j$ elements as follows:
\bea
J_{ij}\eq\f{1}{N}\sum_{\mu=1}^p
x_{i\mu}x_{j\mu}.
\eea
Hereafter {the coefficient $\f{1}{2}$ is included to simplify the discussion below.} 
The method used {in the analysis of the 
minimization of investment 
risk under two constraints is basically similar to 
those in Refs.~\cite{Shinzato-2016-PRE12,Shinzato-2016-PRE11}.
}

In Ref.~\cite{Shinzato-2016-PRE11}, 
the minimal investment risk per asset 
$\ve=\f{1}{N}{\cal H}(\vec{w}^*|X)$, 
the investment concentration 
$q_w=\f{1}{N}(\vec{w}^*)^{\rm T}\vec{w}^*$, and 
the Sharpe ratio $S=\f{R}{\sqrt{2\ve}}$ were derived and shown respectively to be
\bea
\label{eq5}
\ve\eq\f{s^2(\a-1)}{2}\left[
1+\f{(R-m)^2}{\s^2}
\right],\\
\label{eq6}
q_w\eq\f{\a}{\a-1}\left[
1+\f{(R-m)^2}{\s^2}
\right],\\
S\eq\f{R}{\sqrt{s^2(\a-1)\left[
1+\f{(R-m)^2}{\s^2}
\right]}},
\eea
where $\vec{w}^*$ is the portfolio which can minimize the investment risk 
${\cal H}(\vec{w}|X)$, {and thus 
they all depend on} period ratio $\a=p/N\sim O(1)$. 
The Sharpe ratio 
is a criterion defined as 
the ratio of the expected return per asset to 
the square root 
of twice the investment risk per asset.
Note that 
if the investment risk is constant, 
the larger the expected return is, the better the portfolio is;  
and if the expected return is constant, 
the smaller the investment risk is, the better the portfolio is.
{In either case, 
rational investors seek the portfolio which can maximize the Sharpe ratio.}
For {an interpretation} of 
investment concentration, see Ref.\ 
\cite{Shinzato-2016-PRE12}.

{In the above-mentioned previous work, it was} assumed that 
the variance of the return rate of each asset is unique, that is, 
$V[\bar{x}_{i\mu}](=v_i)=s^2$, and 
the hyperparameters of the means $E[\bar{x}_{i\mu}]=r_i$
are independently and identically Gaussian distributed 
with mean $m$ and variance $\s^2$.
As in Ref.~\cite{Shinzato-2016-PRE12}, our aim in this paper is to 
analyze the
minimization of investment risk 
under 
these two constraints for the case that 
the distributions of the hyperparameters of mean $r_i$ and variance $v_i$
are not limited 
{to a specific probability family;
 we here propose an analytical approach based on replica analysis.}

\section{Replica analysis\label{sec3}}
In this section, 
we analyze the 
minimization of investment risk 
under {constraints of budget and expected return} by using a replica analysis technique
which {was} developed previous studies {\cite{Shinzato-2015-PLOS7,Shinzato-2016-PRE12,Shinzato-2011-IEICE,Shinzato-2017-JSTAT2,Shinzato-ve-fixed2016,Shinzato-2016-PRE11}}.
The partition function of this investment system 
at the inverse temperature $\b(>0)$, $Z(R,X,\vec{r})$, {is defined as follows:}
\bea
Z(R,X,\vec{r})\eq
\int_{\vec{w}\in{\cal W}}
d\vec{w}e^{-\b{\cal H}(\vec{w}|X)},
\eea
where 
${\cal W}=\left\{\vec{w}\in{\bf R}^N|\vec{w}^{\rm T}\vec{e}=N,\vec{w}^{\rm T}\vec{r}=NR\right\}$ is the feasible portfolio 
subset space characterized by Eqs. (\ref{eq1}) and (\ref{eq2}),
$\vec{e}=(1,1,\cdots,1)^{\rm T}\in{\bf R}^N$, and 
$\vec{r}=(r_1,r_2,\cdots,r_N)^{\rm T}\in{\bf R}^N$ {are employed}.
Then, using 
\bea
\label{eq9}
\phi
\eq\lim_{N\to\infty}
\f{1}{N}
E\left[\log 
Z(R,X,\vec{r})
\right]
\nn
\eq\lim_{N\to\infty}
\f{1}{N}\lim_{n\to0}\pp{}{n}
\log E
\left[
Z^n(R,X,\vec{r})
\right],
\eea
the minimal investment risk per asset $\ve$ is given by
\bea
\label{eq10}
\ve\eq-\lim_{\b\to\infty}\pp{\phi}{\b},
\eea
where 
the notation $E[f(X,\vec{r})]$ 
means the expectation of 
$f(X,\vec{r})$, in which 
the return rate matrix is $X$ and the vector of hyperparameter of the mean is $\vec{r}$.
Using the ansatz of the replica symmetry solution discussed {in previous studies \cite{Shinzato-2016-PRE11,Shinzato-2016-PRE12,Shinzato-2017-JSTAT2,Shinzato-ve-fixed2016,
Shinzato-2015-PLOS7}}, 
$E
\left[
Z^n(R,X,\vec{r})
\right]$ for 
$n\in{\bf Z}$ and $\phi$ are 
assessed.
Here the replica symmetric solution is 
\bea
\label{eq11-1}
q_{wab}\eq\f{1}{N}\sum_{i=1}^Nw_{ia}w_{ib}\nn
\eq\left\{
\begin{array}{ll}
\chi_w+q_w&a=b\\
q_w&a\ne b
\end{array}
\right.,\\
\label{eq11}
q_{sab}\eq\f{1}{N}\sum_{i=1}^Nv_iw_{ia}w_{ib}\nn
\eq\left\{
\begin{array}{ll}
\chi_s+q_s&a=b\\
q_s&a\ne b
\end{array}
\right.,\\
\tilde{q}_{wab}\eq
\left\{
\begin{array}{ll}
\tilde{\chi}_w-\tilde{q}_w&a=b\\
-\tilde{q}_w&a\ne b
\end{array}
\right.,\\
\label{eq13}
\tilde{q}_{sab}\eq
\left\{
\begin{array}{ll}
\tilde{\chi}_s-\tilde{q}_s&a=b\\
-\tilde{q}_s&a\ne b
\end{array}
\right.,\\
k_a\eq k,\\
\theta_a\eq\theta,
\label{eq16-1}
\eea
where 
$\vec{w}_a=(w_{1a},w_{2a},\cdots,w_{Na})^{\rm T}\in{\bf R}^N,(a,b=1,2,\cdots,n)$,
$\tilde{q}_{wab}$ and $\tilde{q}_{sab}$ are the 
auxiliary variables of 
$q_{wab}$ and $q_{sab}$, respectively, 
$k_a$ is 
the 
auxiliary variable related to  
the budget constraint in \sref{eq1},
and $\theta_a$ is 
the 
auxiliary variable related to  
the expected return constraint in 
\sref{eq2}.
From these settings,
using replica symmetric solution,
\bea
\label{eq17}
\phi
\eq\mathop{\rm Extr}_{\Theta}
\left\{
-\f{\a}{2}\log(1+\b\chi_s)
-\f{\a\b q_s}{2(1+\b\chi_s)}
-k-R\theta
\right.\nn
&&-\f{1}{2}
\left\langle
\log(\tilde{\chi}_w+v\tilde{\chi}_s)
\right\rangle
+\f{1}{2}
\left\langle
\f{\tilde{q}_w+v\tilde{q}_s}
{\tilde{\chi}_w+v\tilde{\chi}_s}
\right\rangle\nn
&&
+\f{1}{2}
\left\langle
\f{(k+r\theta)^2}
{\tilde{\chi}_w+v\tilde{\chi}_s}
\right\rangle
+\f{1}{2}(\chi_w+q_w)(\tilde{\chi}_w-\tilde{q}_w)+\f{q_w\tilde{q}_w}{2}
\nn
&&\left.
+\f{1}{2}(\chi_s+q_s)(\tilde{\chi}_s-\tilde{q}_s)+\f{q_s\tilde{q}_s}{2}
\right\}
\eea
is analyzed, where 
$\a=p/N\sim O(1)$, 
and the notation 
${\rm Extr}_mg(m)$ means 
the extremum of $g(m)$ with respect to $m$.
Furthermore, 
$\Theta=\left\{k,\theta,\chi_w,q_w,\tilde{\chi}_w,\tilde{q}_w,
\chi_s,q_s,\tilde{\chi}_s,\tilde{q}_s
\right\}$ represents the 
set of the order parameters.
The notation 
\bea
\left\langle
f(r,v)
\right\rangle
\eq\lim_{N\to\infty}\f{1}{N}\sum_{i=1}^N
f(r_i,v_i),
\eea
is also used.
Note that the 
deviation  of 
$\phi$ in 
\sref{eq17} is discussed in
appendix \ref{app-1}.

{From the extremum conditions for \sref{eq17}} with respect to these parameters, 
the primal order parameters are as follows: 
\bea
\chi_s\eq\f{1}{\b(\a-1)},\\
q_s\eq\f{\a}{(\a-1)\left\langle v^{-1}\right\rangle}
\left(
1+\f{\left(R-
R_1
\right)^2}
{V_1
}
\right),\\
\chi_w\eq\f{\left\langle v^{-1}\right\rangle}{\b(\a-1)},\\
\label{eq22}q_w\eq\f{1}{\a-1}
\left(
1+\f{\left(R-
R_1
\right)^2}
{V_1
}
\right)
+\f{\left\langle v^{-2}\right\rangle c(R)}{\left\langle v^{-1}\right\rangle^{2}
V_1^2
},
\eea
where 
\bea
\label{eq23}
R_1\eq\f{\left\langle v^{-1}r\right\rangle}
{\left\langle v^{-1}\right\rangle},\\
R_2\eq\f{\left\langle v^{-2}r\right\rangle}
{\left\langle v^{-2}\right\rangle},\\
\label{eq25}
V_1\eq 
\f{\left\langle v^{-1}r^2\right\rangle}
{\left\langle v^{-1}\right\rangle}-
\left(
\f{\left\langle v^{-1}r\right\rangle}
{\left\langle v^{-1}\right\rangle}
\right)^2,\\
V_2\eq\f{\left\langle v^{-2}r^2\right\rangle}
{\left\langle v^{-2}\right\rangle}
-\left(
\f{\left\langle v^{-2}r\right\rangle}
{\left\langle v^{-2}\right\rangle}\right)^2,\\
c(R)\eq V_2
\left(R-R_1\right)^2
+\left(V_1+(R
-R_1)(R_2-R_1)
\right)^2.\nn
\eea
From these results, 
the minimal investment risk per asset $\ve$
is 
derived as follows 
using 
$\ve=-\lim_{\b\to\infty}\pp{\phi}{\b}=\lim_{\b\to\infty}\left(\f{\a\chi_s}{2(1+\b\chi_s)}+\f{\a q_s}{2(1+\b\chi_s)^2}\right)$
in 
\sref{eq10}:
\bea
\label{eq29}
\ve\eq\f{\a-1}{2{\left\langle v^{-1}\right\rangle}}
\left(
1+\f{\left(R-
R_1
\right)^2}
{V_1
}
\right).
\eea
In addition,
Sharpe ratio $S=\f{R}{\sqrt{2\ve}}$ is given by
\bea
\label{eq30}
S\eq\sqrt{\f{\left\langle v^{-1}\right\rangle}{\a-1}}\f{R}{\sqrt{1+\f{(R-R_1)^2}{V_1}}}.
\eea
Note that the investment concentration $q_w$ was
derived in 
\sref{eq22}.
\section{Discussion\label{sec4}}
In this section, 
several properties of the proposed approach will be discussed {in detail}.

\subsection{Comparison with the results derived using the Lagrange multiplier method}
First, 
we will 
derive the minimal investment risk per asset $\ve$
by using the Lagrange multiplier method, and 
compare the results with those of replica analysis.
Here, the Lagrange multiplier $L$ is defined as follows: 
\bea
L\eq\f{1}{2}\vec{w}^{\rm T}J\vec{w}+k(N-\vec{w}^{\rm T}\vec{e})+\theta
(NR-\vec{w}^{\rm T}\vec{r}).
\eea
Then the optimal portfolio $\vec{w}^*$ is 
obtained by solving $\pp{L}{\vec{w}}=0,\pp{L}{k}=\pp{L}{\theta}=0$ to give 
\bea
\vec{w}^*\eq kJ^{-1}\vec{e}+\theta J^{-1}\vec{r},\\
\left(
\begin{array}{c}
k\\
\theta
\end{array}
\right)
\eq
\f{1}{D}
\left(
\begin{array}{cc}
\f{\vec{r}^{\rm T}J^{-1}\vec{r}}{N}&
-\f{\vec{r}^{\rm T}J^{-1}\vec{e}}{N}\\
-\f{\vec{r}^{\rm T}J^{-1}\vec{e}}{N}&
\f{\vec{e}^{\rm T}J^{-1}\vec{e}}{N}
\end{array}
\right)
\left(
\begin{array}{c}
1\\
R
\end{array}
\right),
\eea
where 
\bea
D\eq
\left(\f{\vec{e}^{\rm T}J^{-1}\vec{e}}{N}\right)^2
\left[
\f{\vec{r}^{\rm T}J^{-1}\vec{r}}{\vec{e}^{\rm T}J^{-1}\vec{e}}
-\left(
\f{\vec{r}^{\rm T}J^{-1}\vec{e}}{\vec{e}^{\rm T}J^{-1}\vec{e}}
\right)^2
\right].
\eea
Thus, 
from the relation 
$\ve=\f{1}{2N}\vec{w}^{\rm T}J\vec{w}=\f{k+R\theta}{2}$, 
the minimal investment risk per asset is
\bea
\label{eq35}
\ve\eq\f{N}{2\vec{e}^{\rm T}J^{-1}\vec{e}}
\left\{
1+\f{\left(R-\f{\vec{r}^{\rm T}J^{-1}\vec{e}}{\vec{e}^{\rm T}J^{-1}\vec{e}}\right)^2}{
\f{\vec{r}^{\rm T}J^{-1}\vec{r}}{\vec{e}^{\rm T}J^{-1}\vec{e}}
-
\left(\f{\vec{r}^{\rm T}J^{-1}\vec{e}}{\vec{e}^{\rm T}J^{-1}\vec{e}}\right)^2
}
\right\}.
\eea
Moreover, 
by the argument in appendix \ref{app-a} 
(\sref{eq-a10} to \sref{eq-a12}), in the limit of a large number of assets $N$,
$\f{1}{N}\vec{e}^{\rm T}J^{-1}\vec{e}=\f{\left\langle v^{-1}\right\rangle}{\a-1}$,
$\f{1}{N}\vec{r}^{\rm T}J^{-1}\vec{e}=\f{\left\langle v^{-1}r\right\rangle}{\a-1}$, and 
$\f{1}{N}\vec{r}^{\rm T}J^{-1}\vec{r}=\f{\left\langle v^{-1}r^2\right\rangle}{\a-1}$
 are {obtained briefly}. 
We substitute {these} into \sref{eq35} to obtain
\bea
\ve\eq\f{\a-1}{2{\left\langle v^{-1}\right\rangle}}
\left(
1+\f{\left(R-
R_1
\right)^2}
{V_1
}
\right)
\eea
in terms of 
$R_1$ in \sref{eq23} and $V_1$ in \sref{eq25}.
Thus, the result using the Lagrange multiple method is 
identical to that using replica analysis {in} \sref{eq29}. 

\subsection{Dual optimization problem\label{sec4.2}}
Next, we will 
discuss the dual problem of the minimization of investment risk problem under constraints of budget and expected return, 
which is equivalent to 
the maximization of expected return problem under
constraints of budget and investment risk.
From an argument made in previous work \cite{Shinzato-ve-fixed2016,Shinzato-2016-PRE11},
the maximum and minimum of the expected return per asset 
$R=\f{1}{N}\sum_{i=1}^Nr_iw_i$ can be written as follows:
\bea
R^{\max}\eq
\lim_{N\to\infty}
\mathop{\max}_{\vec{w}\in{\cal W}'}
\left\{\f{1}{N}\sum_{i=1}^Nr_iw_i\right\},\\
R^{\min}\eq
\lim_{N\to\infty}
\mathop{\min}_{\vec{w}\in{\cal W}'}
\left\{\f{1}{N}\sum_{i=1}^Nr_iw_i\right\}.
\eea
That is, we can define two dual problems systematically using 
the feasible portfolio subset space characterized 
by the constraints of budget and investment risk, which is written as follows:
\bea
{\cal W}'\eq
\left\{
\vec{w}\in{\bf R}^N
\left|\vec{w}^{\rm T}\vec{e}=N,
\f{1}{2}\vec{w}^{\rm T}J\vec{w}=N\ve
\right.
\right\}.
\eea 
As shown in previous work {\cite{Shinzato-2016-PRE11}},  
it is also easy to solve this dual problem by using replica analysis {(see also appendix \ref{app-1}). 
Specifically, we can find the upper and lower bounds on expected return by using \sref{eq29} as follows:}
\bea
\label{eq39-1}
R^{\max}\eq R_1+\sqrt{V_1\left(\f{2\left\langle v^{-1}\right\rangle}{\a-1}\ve-1\right)},\\
\label{eq40-1}
R^{\min}\eq R_1-\sqrt{V_1\left(\f{2\left\langle v^{-1}\right\rangle}{\a-1}\ve-1\right)}.
\eea

\subsection{Comparison with the results under only the budget constraint}
We will ascertain whether 
the minimization of investment risk problem 
under only the budget constraint 
analyzed in previous work \cite{Shinzato-2016-PRE12} is included in the analytical results of the present paper.
In the previous work, 
the variances of the return rates of the assets 
were not identical. That is, 
since $V[\bar{x}_{i\mu}](=v_i)=s_i$, 
\bea
\left\langle v^{-1}
\right\rangle\eq\lim_{N\to\infty}\f{1}{N}\sum_{i=1}^N\f{1}{s_i},
\eea
and the right-hand side is rewritten as  
$\left\langle s^{-1}\right\rangle$. 
Then the minimal investment risk per asset of the minimization portfolio problem under the 
budget constraint only, 
$\ve_0$, can be described as $\ve_0=\f{\a-1}{2\left\langle s^{-1}\right\rangle}$, which is 
the first term in \sref{eq29}.
From this, 
the second term in \sref{eq29}, 
$\f{\a-1}{2\left\langle s^{-1}\right\rangle}\f{(R-R_1)^2}{V_1}$, is 
related to the expected return constraint.
Moreover, 
using $E[\bar{x}_{i\mu}](=r_i)=R$, 
since the budget constraint in \sref{eq1} 
can be {equivalent to}
the expected return constraint in 
\sref{eq2}, 
the minimal investment risk per asset $\ve$ takes 
its minimum; that is,  
from $R=R_1$,
the second term in \sref{eq29}, $\f{\a-1}{2\left\langle s^{-1}\right\rangle}\f{(R-R_1)^2}{V_1}$, is 0 and 
if $V_1=V_2\to0$, then $c(R)/V_1^2\to1$, implying 
\bea
\ve\eq\f{\a-1}{2\left\langle s^{-1}\right\rangle},\\
q_w\eq\f{1}{\a-1}+\f{\left\langle s^{-2}\right\rangle}{\left\langle s^{-1}\right\rangle^2}.
\eea
Namely, 
the result obtained in previous work (\sref{eq5} and \sref{eq6}) is included in the present analysis 
(recall that $\left\langle s^{-2}\right\rangle=\lim_{N\to\infty}\f{1}{N}\sum_{i=1}^N\f{1}{s_i^2}$).

\subsection{\label{sec4.4}
Comparison with the results under constraints of budget and expected return}
Let us now clarify 
that 
the analytical results of 
the 
minimization of 
investment risk problem 
under
constraints of budget and expected return previously reported \cite{Shinzato-2016-PRE11}
are replicated 
by our proposed approach.
Here, 
the variance of return rate of each asset is a constant, that is, $V[\bar{x}_{i\mu}](=v_i)=s^2$,
and $E[\bar{x}_{i\mu}]=r_i$ are independently 
and identically Gaussian distributed with 
mean $m$ and 
variance $\s^2$.
Then,
$\left\langle v^{-1}\right\rangle=s^{-2}, R_1=R_2=m,V_1=V_2=\s^2$, and $c(R)=\s^2(R-m)^2+\s^4$, 
which gives 
\bea
\label{eq43}
\ve\eq\f{s^2(\a-1)}{2}\left(1+\f{(R-m)^2}{\s^2}\right),\\
\label{eq44-1}
q_w\eq\f{\a}{\a-1}\left(1+\f{(R-m)^2}{\s^2}\right).
\eea
Thus, our results here agree with those in previous work. In addition, 
when $r_i$ and $v_i$ are uncorrelated with each other,
$R_1=R_2=\left\langle r\right\rangle=m$ and 
$V_1=V_2=\left\langle r^2\right\rangle
-\left\langle r\right\rangle^2
=\s^2$, which has no effect on 
$q_w$
in \sref{eq44-1}.  
Note that, using relation $s^2=\left\langle v^{-1}\right\rangle^{-1}$, \sref{eq43} takes the following form:
\bea
\ve
\eq\f{\a-1}{2\left\langle v^{-1}\right\rangle}\left(1+\f{(R-m)^2}{\s^2}\right).
\eea

\subsection{Pythagorean theorem of the Sharpe ratio}
{As an innovative highlight of the proposed approach},
let us discuss the 
macroscopic relationship of 
Sharpe ratio $S=\f{R}{\sqrt{2\ve}}$.
From \sref{eq30},
the maximal Sharpe ratio 
$S(R^*)$ 
occurs 
at 
$R=R^*=\f{R_1^2+V_1}{R_1}=\f{\left\langle v^{-1}r^2\right\rangle}{\left\langle v^{-1}r\right\rangle}$, with
\bea
\label{eq45}
S(R^*)\eq\sqrt{\f{\left\langle v^{-1}\right\rangle}{\a-1}}\sqrt{R_1^2+V_1}.
\eea
Further, 
the case of having the budget constraint only ($R=R_1$) and 
the case that the return coefficient $R$ is set at infinity, 
\bea
S(R_1)\eq\sqrt{\f{\left\langle v^{-1}\right\rangle}{\a-1}}R_1,\\
S(\infty)\eq\sqrt{\f{\left\langle v^{-1}\right\rangle}{\a-1}}\sqrt{V_1},
\eea
can also be obtained. 
Using these results, 
the following relation can be proved, 
which we call 
the Pythagorean theorem of the Sharpe ratio:
\bea
\label{eq50}
S^2(R^*)
\eq S^2(R_1)+S^2(\infty).
\eea

{Equation 
(\ref{eq50})}
 is interpreted as follows.
Using \sref{eq29}, since 
the minimal investment risk per asset $\ve$ is 
a quadratic function of $R$,
$R=R_1$ is the return coefficient which can minimize the 
minimal investment risk, and 
$R\to\infty$ is 
the return coefficient 
which 
can maximize the 
minimal investment risk, 
for convenience sake,
it can be interpreted that 
the square sum of Sharpe ratios at the two extremes
$S^2(R_1)+S^2(\infty)$ is 
consistent  with the square of the maximal Sharpe ratio $S^2(R^*)$.
Note that 
the strong theorem in \sref{eq50} holds 
at for any $\a>1$ and 
arbitrary distributions of the hyperparameters $E[\bar{x}_{i\mu}]=r_i$ and 
{$V[\bar{x}_{i\mu}]=v_i$.}
Moreover, 
this theorem 
is distinct from the Pythagorean theorem of a rectangular triangle; 
though the geometrical interpretation is not yet clear, 
this theorem 
could imply 
new macroscopic relations (similar to thermodynamic relations) related to 
mathematical finance.

\subsection{Maximization of Sharpe ratio}
Next, we will discuss 
the maximal Sharpe ratio 
without using 
replica analysis. 
By using Eqs. (\ref{eq2}) and (\ref{eq3}),
Sharpe ratio $S=\f{R}{\sqrt{2\ve}}$ 
is generalized to 
$S=\f{\f{1}{N}\vec{r}^{\rm T}\vec{w}}{\sqrt{\f{1}{N}\vec{w}^{\rm T}J\vec{w}}}$,
based on Cauchy--Schwarz inequality $\left|\vec{a}^{\rm T}\vec{b}\right|\le\sqrt{\vec{a}^{\rm T}\vec{a}}\sqrt{\vec{b}^{\rm T}\vec{b}}$,
since $\f{\vec{a}^{\rm T}\vec{b}}{\sqrt{\vec{b}^{\rm T}\vec{b}}}$ takes 
a maximum value $\sqrt{\vec{a}^{\rm T}\vec{a}}$
at 
$\vec{b}=K\vec{a},(K>0)$.
Then the maximal Sharpe ratio $S(R^*)$ is
\bea
\label{eq49}
S(R^*)\eq\sqrt{\f{1}{N}\vec{r}^{\rm T}J^{-1}\vec{r}},
\eea
where 
$\vec{a}=J^{-\f{1}{2}}\vec{r}$ and 
$\vec{b}=J^{\f{1}{2}}\vec{w}$ have already been employed.
Furthermore, 
from $\vec{b}=K\vec{a}$,
{$\vec{w}=KJ^{-1}\vec{r}$,
when the coefficient $K$} is $K=\f{N}{\vec{r}^{\rm T}J^{-1}\vec{e}}$, 
\sref{eq1} is 
satisfied.
From this, 
the expected return which can maximize the Sharpe ratio, 
$R^*=\f{1}{N}\vec{r}^{\rm T}\vec{w}$, is given by
\bea
R^*\eq\f{\vec{r}^{\rm T}J^{-1}\vec{r}}
{\vec{r}^{\rm T}J^{-1}\vec{e}}.
\eea
From an argument in appendix \ref{app-a},
$R^*=\f{\left\langle v^{-1}r^2\right\rangle}{\left\langle v^{-1}r\right\rangle}$,
which is consistent with the result in the previous subsection.
Further, 
in a similar way, 
from $\f{1}{N}\vec{r}^{\rm T}J^{-1}\vec{r}=\f{\left\langle v^{-1}r^2\right\rangle}{\a-1}$,
$S(R^*)$ in \sref{eq49} is given by
\bea
S(R^*)
\eq\sqrt{\f{\left\langle v^{-1}\right\rangle}{\a-1}}
\sqrt{\f{\left\langle v^{-1}r^2\right\rangle}
{\left\langle v^{-1}r\right\rangle}
}
\sqrt{\f{\left\langle v^{-1}r\right\rangle}
{\left\langle v^{-1}\right\rangle}
}
\nn
\eq
\sqrt{\f{\left\langle v^{-1}\right\rangle}{\a-1}}
\sqrt{\f{R_1^2+V_1}{R_1}}
\sqrt{R_1},
\eea
where 
$\f{\left\langle v^{-1}r^2\right\rangle}{\left\langle v^{-1}r\right\rangle}=\f{R_1^2+V_1}{R_1}$
and 
$\f{\left\langle v^{-1}r\right\rangle}{\left\langle v^{-1}\right\rangle}=R_1$
have already been applied. Thus this result agrees  
with \sref{eq45}.
\subsection{Comparison with the result based on operations research}

Finally, 
we should 
compare the results derived using 
the standard approach 
in operations research \cite{Markowitz1952,Markowitz1959,Bodie,Luenberger,Prigent,Ang,Francis,Elton} with 
those derived {from our} replica analysis.
Firstly, 
following 
{the standard analytical procedure,} 
the expected investment risk $E[{\cal H}(\vec{w}|X)]$ is estimated as follows:
\bea
\label{eq54}
E[{\cal H}(\vec{w}|X)]\eq\f{\a}{2}\sum_{i=1}^Nv_iw_i^2.
\eea
Next, the portfolio which can minimize the expected investment risk $E[{\cal H}(\vec{w}|X)]$ under the budget constraint in
\sref{eq1} and the 
expected return constraint in \sref{eq2}, 
$\vec{w}^{\rm OR}=(w_1^{\rm OR},\cdots,w_N^{\rm OR})^{\rm T}
=\arg\mathop{\min}_{\vec{w}\in{\cal W}}
E[{\cal H}(\vec{w}|X)]
\in{\bf R}^N$, can be determined, giving the
following minimal expected investment risk per asset:
\bea
\ve^{\rm OR}\eq\lim_{N\to\infty}\f{1}{N}E[{\cal H}(\vec{w}^{\rm OR}|X)]\nn
\eq
\f{\a}{2{\left\langle v^{-1}\right\rangle}}
\left(
1+\f{\left(R-
R_1
\right)^2}
{V_1
}
\right),
\eea
Therefore, 
the opportunity loss of the portfolio which is provided by the approach 
of operations research, $\vec{w}^{\rm OR}$, 
that is, 
$\kappa=\f{\ve^{\rm OR}}{\ve}$, is calculated as follows:
\bea
\label{eq56}
\kappa
\eq\f{\a}{\a-1}.
\eea
Namely, the portfolio which can minimize the expected investment risk 
$E[{\cal H}(\vec{w}|X)]$ (but not the investment risk ${\cal H}(\vec{w}|X)$),
$\vec{w}^{\rm OR}$, 
does not always minimize 
${\cal H}(\vec{w}|X)$.
From this, 
it is clarified that 
the standard analytical procedure
provides 
a 
portfolio $\vec{w}^{\rm OR}$ which 
does not {consider}
the diversification of risk, unlike  
the optimal portfolio obtained by our proposed approach 
(see appendix \ref{app-b} for details).
Notice that since the opportunity loss in \sref{eq56}
depends on $\a$ and not on 
the distributions of the hyperparameters $E[\bar{x}_{i\mu}]=r_i$
and $V[\bar{x}_{i\mu}]=v_i$, 
this macroscopic relation  between risks 
holds in a similar fashion to the Pythagorean theorem of the Sharpe ratio given in 
\sref{eq50}.

Similarly, {the investment concentration of the standard analytical procedure} 
$q_w^{\rm OR}=\lim_{N\to\infty}\f{1}{N}\sum_{i=1}^N(w_i^{\rm OR})^2$ is 
evaluated as follows:
\bea
\label{eq40}
q_w^{\rm OR}
\eq
\f{\left\langle v^{-2}\right\rangle c(R)}{\left\langle v^{-1}\right\rangle^{2}
V_1^2
}.
\eea
That is, $q_w^{\rm OR}$ is the same as the second term in 
\sref{eq22}.
In addition, when $\a$ is close to 1,
in general, 
rational 
investors 
tend to
 invest intensively in assets of comparatively small risk. 
If 
the reference return rate is set as $X=\left\{\f{x_{i\mu}}{\sqrt{N}}\right\}\in{\bf R}^{N\times p}$ 
\cite{Shinzato-2015-PLOS7,Shinzato-2016-PRE11,Shinzato-2016-PRE12},
such investment behavior
is well known to cause the investment concentration $q_w$ 
to be large. Namely, 
$q_w$ in \sref{eq22} is 
successful at expressing the 
optimal investment behavior and 
$q_w^{\rm OR}$
in \sref{eq40} 
fails to 
take into account the optimal investment strategy.
Thus, 
the portfolio which can 
minimize the expected investment risk $\vec{w}^{\rm OR}=\arg\mathop{\min}_{\vec{w}\in{\cal W}}E[{\cal H}(\vec{w}|X)]$
unfortunately fails to include some important investment properties that are 
possessed by $\vec{w}^*=\arg\mathop{\min}_{\vec{w}\in{\cal W}}{\cal H}(\vec{w}|X)$.

In addition, 
other sorts of risk than 
the minimal 
investment risk ${\cal H}(\vec{w}^*|X)$ and 
the minimal expected investment 
risk $E[{\cal H}(\vec{w}^{\rm OR}|X)]$ can be considered.
For instance, 
one can substitute 
the optimal portfolio $\vec{w}^*=\arg\mathop{\min}_{\vec{w}\in{\cal W}}{\cal H}(\vec{w}|X)$
into the expected investment 
risk $E[{\cal H}(\vec{w}|X)]$ in 
\sref{eq54} to obtain 
the expected investment risk per asset of 
the optimal portfolio $\vec{w}^*$, that is, 
$\ve'=\lim_{N\to\infty}
\lim_{\vec{w}\to\vec{w}^*}
\f{1}{N}E[{\cal H}(\vec{w}|X)]$ is estimated as follows;
\bea
\ve'
\eq\f{\a}{2}\lim_{N\to\infty}
\f{1}{N}\sum_{i=1}^Nv_i(w_i^*)^2\nn
\eq\f{\a}{2}\left(\chi_s+q_s\right),
\eea
where $q_{saa}=\chi_s+q_s$
in \sref{eq11} is used. 
If $\b\to\infty$, $\chi_s\to0$ is obtained. That is, 
$q_s$ defined in {our} replica analysis corresponds to $\ve'$,
the expected 
investment risk per asset of 
the optimal portfolio $\vec{w}^*$. Moreover, 
the opportunity loss of 
$\ve'$ with respect to 
the minimal investment risk per asset $\ve$, that
is,  $\kappa'=\f{\ve'}{\ve}$, is as follows:
\bea
\label{eq59-1}
\kappa'\eq\left(\f{\a}{\a-1}\right)^2.
\eea
Notice that 
the opportunity loss in 
\sref{eq59-1}, 
$\kappa'$, 
depends on 
$\a$ and not on 
the distributions of 
the hyperparameters 
$E[\bar{x}_{i\mu}]=r_i$ and 
$V[\bar{x}_{i\mu}]=v_i$ 
in a similar way as the opportunity loss $\kappa$ in \sref{eq56} {(see Table \ref{tab2})}.

\begin{table}[t]
{	\centering
{
\caption{Comparison of typical risks per asset. Note that 
the upper left entry $\ve$ and lower right entry $\ve^{\rm OR}$ define 
the opportunity loss $\kappa$, the upper left entry $\ve$ and lower left entry $\ve'$
define the opportunity loss $\kappa'$,
and the upper right entry is consistent with the lower right entry, 
the expectation of ${\cal H}(\vec{w}^{\rm OR}|X)$. 
\label{tab2}}
}
	\begin{tabular}{|c|c|c|}\hline
&$\vec{w}^*$&$\vec{w}^{\rm OR}$\\
\hline
${\cal H}(\vec{w}|X)$&$\ve$&$\ve^{\rm OR}$\\
\hline
$E[{\cal H}(\vec{w}|X)]$&$\ve'$&$\ve^{\rm OR}$\\
\hline
\end{tabular}
}
\end{table}

\section{Numerical experiments\label{sec5}}
In this section, 
we 
verify the effectiveness of 
our proposed method 
by using a numerical experiment.
From the discussion in subsection 
\ref{sec4.4}, 
if $r_i$ and $v_i$ are independently distributed {with respect to each other}, 
the results using the proposed approach are consistent with 
those using our previously reported approach  \cite{Shinzato-2016-PRE11}; 
therefore, we will next consider the case that $r_i$ and $v_i$ are correlated with each other.
For instance, 
recalling that $r_i=E[\bar{x}_{i\mu}]$ and 
$v_i=E[\bar{x}_{i\mu}^2]-(E[\bar{x}_{i\mu}])^2$, 
we assume that $E[\bar{x}_{i\mu}^2]$ is proportional to $r_i^2$,
that is, $E[\bar{x}_{i\mu}^2]=(h_i+1)r_i^2$. 
Here, $h_i$ is a random coefficient 
to simplify the description in \sref{eq59},
and variance $v_i$ is described using the square of the hyperparameter of the mean, $r_i^2$, and 
$h_i$ as follows:
\bea
\label{eq59}
v_i\eq h_i r_i^2.
\eea

Then $r_i$ and $h_i$ are independently distributed with 
Pareto distributions within the bounded interval $(l_r\le r_i\le u_r,l_h\le h_i\le u_h)$ and these 
probability density functions (which we call the bounded Pareto distributions with {the powers} $c_r$ and $c_h$, respectively) are defined as follows \cite{Aban}:
\bea
\label{eq60}
f_r(r_i)\eq
\left\{
\begin{array}{ll}
\f{1-c_r}{u_r^{1-c_r}-l_r^{1-c_r}
}r_i^{-c_r}&l_r\le r_i\le u_r\\
0&\text{otherwise}
\end{array}
\right.,\\
\label{eq61}
f_h(h_i)\eq
\left\{
\begin{array}{ll}
\f{1-c_h}{u_h^{1-c_h}-l_h^{1-c_h}
}h_i^{-c_h}&l_h\le h_i\le u_h\\
0&\text{otherwise}
\end{array}
\right..
\eea
That is, the parameters of the density functions $f_r(r_i)$ and $f_h(h_i)$ of $r_i$ and $h_i$
are 
$(l_r,u_r,c_r)$ and 
$(l_h,u_h,c_h)$, respectively, where
$l_r,l_h,c_r,c_h>0$ is assumed.
In addition, 
in the case that $\l,\l'$ are independently and identically uniformly distributed over the 
interval $0\le\l,\l'\le1$, 
$r_i,h_i$ are assigned as 
$r_i=(\l u_r^{1-c_r}+(1-\l)l_r^{1-c_r})^{\f{1}{1-c_r}}$
 and 
$h_i=(\l' u_h^{1-c_h}+(1-\l')l_h^{1-c_h})^{\f{1}{1-c_h}}$, respectively. 
That is, they are drawn from the probability density functions in \sref{eq60} and \sref{eq61}.

We can derive numerically 
the minimal investment risk per asset $\ve$ using the following steps:
\begin{description}
\item[Step 1]
Assign $r_i$ and $h_i$ independently to the bounded Pareto distributions in Eqs. (\ref{eq60}) and (\ref{eq61}); 
{in addition to setting the hyperparameter of mean $r_i$, 
we can prepare the hyperparameter of variance 
$v_i(=h_ir_i^2)$. }
\item[Step 2]
Draw the return rate of asset $i$ at period $\mu$, $\bar{x}_{i\mu}$, 
from a probability distribution such that 
$E[\bar{x}_{i\mu}]=r_i$ and $V[\bar{x}_{i\mu}]=v_i$.
Calculate the modified 
return rate $x_{i\mu}=\bar{x}_{i\mu}-r_i$ to construct the 
return rate matrix $X=\left\{\f{x_{i\mu}}{\sqrt{N}}\right\}\in{\bf R}^{N\times p}$.
\item[Step 3] Calculate $J=XX^{\rm T}\in{\bf R}^{N\times N}$ and 
the inverse matrix $J^{-1}$.
\item[Step 4] Evaluate $\f{1}{N}\vec{e}^{\rm T}J^{-1}\vec{e},
\f{1}{N}\vec{r}^{\rm T}J^{-1}\vec{e}$, and
$\f{1}{N}\vec{r}^{\rm T}J^{-1}\vec{r}$.
\item[Step 5]
Evaluate the minimal investment risk per asset $\ve$ by using \sref{eq35}.
\end{description}

In order to assess the typical behavior of 
the minimal investment risk per asset using this procedure, 
$M$ trial experiments are performed.
Specifically, we construct $M$ return rate matrices 
$X^m=\left\{\f{x_{i\mu}^m}{\sqrt{N}}\right\}\in{\bf R}^{N\times p}, (m=1,2,\cdots,M)$, 
$M$ vectors of the hyperparameters of the means of the assets $\vec{r}^m=(r_1^m,r_2^m,\cdots,r_N^m)^{\rm T}\in{\bf R}^N$,
and 
$M$ vectors of the hyperparameters of the variances of the assets 
$\vec{v}^m=(v_1^m,v_2^m,\cdots,v_N^m)^{\rm T}\in{\bf R}^N$ in 
Steps 1 and 2, and determine 
the minimal investment risk per asset at each trial $\ve^m$ in Steps 3 to 5.
The expectation of the minimal 
investment risk per asset ${\ve}$ is then estimated 
as follows:
\bea
{\ve}
\eq\f{1}{M}\sum_{m=1}^M
\ve^m.
\eea
In a similar way, the investment concentration $q_w$
and Sharpe ratio $S$ are also evaluated using 
the above-mentioned steps and 
we compare the 
results with those derived using replica analysis.

\begin{figure}[t] 
\begin{center}
{
\includegraphics[width=0.8\hsize,angle=0]{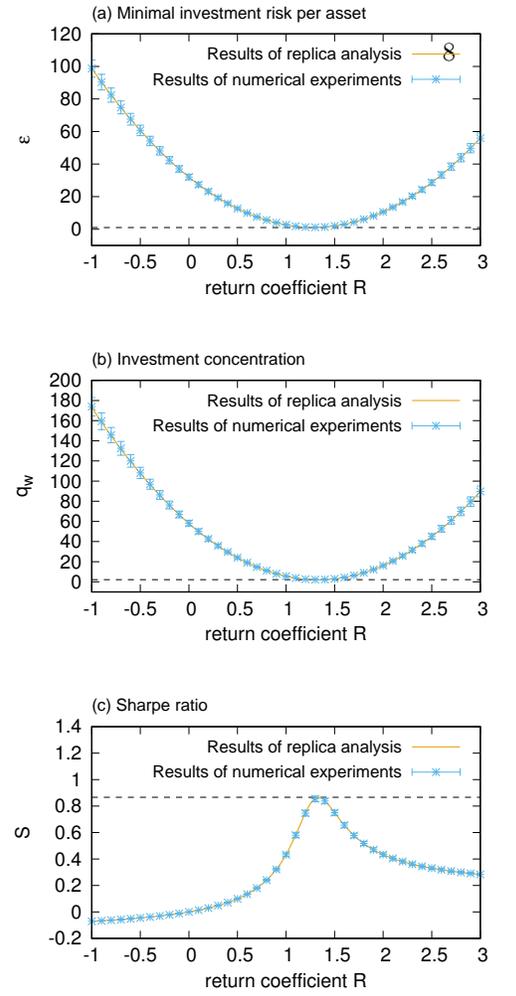}
\caption{
\label{Fig1}
Results of the replica analysis and the numerical experiments ($\a=p/N=2$).
The horizontal axis indicates the return coefficient $R$, and the vertical axes show 
(a) the minimal investment risk per asset $\ve$, (b) the investment 
 concentration $q_w$, and (c) the Sharpe ratio $S$. 
The solid (orange) lines indicate the results of 
the replica analysis for (a) 
 \sref{eq29}, (b) \sref{eq22}, and (c) \sref{eq30}. The (blue) asterisks with 
 error bars indicate the results of the numerical simulation, and 
the  dashed (black) lines indicate the results for (a) $\ve_0=\f{\a-1}{2\left\langle v^{-1}\right\rangle}$, (b) $\f{1}{\a-1}+\f{\left\langle v^{-2}\right\rangle}{\left\langle v^{-1}\right\rangle^2}$, and 
 (c) $S(R^*)=\sqrt{\f{\left\langle v^{-1}r^2\right\rangle}{\a-1}}$. 
}
}
\end{center}
\end{figure}

In this experiment, we use the following settings: $(l_r,u_r,c_r)=(l_h,u_h,c_h)=
(1,2,2)$, number of assets $N=1000$,
number of periods $p=2000$ (that is, as $\a=p/N=2$),
and number of trials $M=100$.
For these numerical settings,
we assess the minimal investment risk per asset, investment concentration, and 
Sharpe ratio, as shown in Fig. \ref{Fig1}. 
From these figures, 
the results are obviously consistent with other. That is, 
these comparisons validate the applicability
of our proposed methodology based on replica analysis.

\section{Conclusion and future work\label{sec6}}
To refine the portfolio optimization problem discussed in previous work \cite{Shinzato-2016-PRE11}, 
which was under constraints of budget and expected return for the case that the variance of the return rate of each asset is unique and 
the hyperparameters of the means of the assets are 
independently and identically Gaussian distributed, 
in the present study we consider
the portfolio 
optimization problem 
under these
two constraints 
for the case that 
the hyperparameters of 
the means and variances of the
assets have 
arbitrary distributions (although, so as to verify our proposed method,
the distributions of hyperparameters were limited in numerical simulations).
Using replica analysis,
the minimal investment risk per asset,
investment concentration, and Sharpe 
ratio of the above-explained optimization problem 
were analytically 
derived.
Moreover, 
by comparing 
the results obtained in previous work, those 
derived using the Lagrange multiplier method, 
and our numerical results,  
the applicability 
of our proposed approach based on 
replica analysis 
was validated.
In addition, 
relations between 
macroscopic variables which are represented by the Pythagorean theorem of the Sharpe ratio 
in \sref{eq50} and 
the two opportunity losses in 
Eqs. (\ref{eq56}) and (\ref{eq59-1}) were derived.
Furthermore, it was shown that  the 
portfolio which is discussed in operations research and which can minimize the expected 
investment risk (which is not the same as the investment risk itself) is not 
always consistent with 
the optimal portfolio
which can minimize the 
investment risk. Since 
the above opportunity losses are larger than 1, from 
the argument in this paper,
as the unfortunate 
consequence, 
it is validated that 
the approach which should {be based on an ill-developed philosophy} 
is not possible to 
attain the optimal asset management which is expected by the rational investors. 
While, fortunately, {interdisciplinary research fields have provided 
step-by-step richer 
knowledge and novel insight for optimal investing }
to rational investors using the analytical approaches {well-developed} in statistical mechanical informatics, we should continue ongoing work to further explore the undeveloped frontier 
in order to develop 
an approach which can derive the optimal 
investment management which meets the expectations of investors.

As future work, 
although the present paper does not discuss mathematically our obtained relation
 between the macroscopic variables sufficiently,
in order to increase the sophistication of 
the body of knowledge of mathematical finance, 
we need to provide a geometrical interpretation of 
the Pythagorean theorem of the Sharpe ratio.
Moreover, 
we also need to 
{determine} additional relations between the macroscopic variables besides the Pythagorean theorem of the Sharpe ratio in \sref{eq50}
and the opportunity losses in (\ref{eq56}) and (\ref{eq59-1}).

\section*{Acknowledgments}
The author is grateful for valuable discussions with K.
Kobayashi, D. Tada, and H. Yamamoto. This work was supported in part by
Grant-in-Aid No. 15K20999; the President Project for Young
Scientists at Akita Prefectural University; Research Project No.
50 of the National Institute of Informatics, Japan; Research
Project No. 5 of the Japan Institute of Life Insurance; Research
Project of the Institute of Economic Research Foundation at
Kyoto University; Research Project No. 1414 of the Zengin
Foundation for Studies in Economics and Finance; Research
Project No. 2068 of the Institute of Statistical Mathematics;
Research Project No. 2 of the Kampo Foundation; and Research
Project of the Mitsubishi UFJ Trust Scholarship Foundation.

\appendix

\section{Replica calculation\label{app-1}}
In this appendix, 
we explain replica analysis in the main context of interest in this paper.
{The same as} in previous work \cite{Shinzato-2015-PLOS7,Shinzato-ve-fixed2016,
Shinzato-2017-JSTAT2,Shinzato-2016-PRE11,Shinzato-2016-PRE12}, 
$E\left[Z^n(R,X,\vec{r})\right],(n\in{\bf Z})$ is 
described as follows:
\bea
&&E\left[Z^n(R,X,\vec{r})\right]\nn
\eq
\mathop{\rm Extr}_{\vec{k},\vec{\theta}}
\f{1}{(2\pi)^{\f{Nn}{2}+pn}}
\area \prod_{a}
d\vec{w}_ad\vec{u}_a
d\vec{z}_a
\nn
&&
E
\left[
\exp
\left(-\f{\b}{2}\sum_{\mu,a}z_{\mu a}^2
\right.
\right.
+\sum_{a}k_a\left(\sum_{i}w_{ia}-N\right)
\nn
&&
+\sum_{a}\theta_a\left(\sum_{i}r_iw_{ia}-NR\right)
\nn
&&
\left.\left.
+i\sum_{\mu,a}u_{\mu a}
\left(z_{\mu a}-\f{1}{\sqrt{N}}\sum_{i}w_{ia}x_{i\mu}\right)
\right)\right],
\eea
where here 
for convenience, 
$\sum_{i}$ indicates $\sum_{i=1}^N$, 
$\sum_{\mu}$ represents $\sum_{\mu=1}^p$, 
$\sum_{a}$ is $\sum_{a=1}^n$, and 
$\prod_{a}$ means 
$\prod_{a=1}^n$. 
Moreover, 
$\vec{w}_a=(w_{1a},w_{2a},\cdots,w_{Na})^{\rm T}\in{\bf R}^N,(a,b=1,2,\cdots,n)$, 
$\vec{u}_a=(u_{1a},u_{2a},\cdots,u_{pa})^{\rm T}\in{\bf R}^p$, 
$\vec{z}_a=(z_{1a},z_{2a},\cdots,z_{pa})^{\rm T}\in{\bf R}^p$, 
$\vec{k}=(k_1,k_2,\cdots,k_n)^{\rm T}\in{\bf R}^n$, and 
$\vec{\theta}=(\theta_1,\theta_2,\cdots,\theta_n)^{\rm T}\in{\bf R}^n$.
Further, the integral $g(\vec{w}_a)$ over
the feasible portfolio subset space (that is, satisfying the budget constraint in \sref{eq1} and the
expected return constraint in \sref{eq2}), ${\cal W}$, is 
approximated  as follows:
\bea
&&\int_{\vec{w}_a\in{\cal W}}d\vec{w}_ag(\vec{w}_a)\nn
\eq\mathop{\rm Extr}_{k_a,\theta_a}
\f{1}{(2\pi)^{\f{N}{2}}}
\area d\vec{w}_ag(\vec{w}_a)
\nn
&&
\exp\left(
k_a\left(\sum_{i}w_{ia}-N\right)
+\theta_a\left(\sum_{i}r_iw_{ia}-N\right)
\right).\nn
\eea
Next, we can assess each part of the integral step by step as follows:
\bea
&&\log E\left[
\exp\left(
-\f{i}{\sqrt{N}}
\sum_{i,\mu}x_{i\mu}\sum_{a}u_{\mu a}w_{ia}
\right)
\right]\nn
\eq
\mathop{\rm Extr}_{Q_w,\tilde{Q}_w,Q_s,\tilde{Q}_s}
\left\{
-\f{1}{2}\sum_{\mu,a,b}q_{sab}u_{\mu a}u_{\mu b}
\right.\nn
&&
-\f{1}{2}\sum_{a,b}\tilde{q}_{wab}
\left(\sum_{i}w_{ia}w_{ib}-Nq_{wab}\right)\nn
&&\left.
-\f{1}{2}\sum_{a,b}\tilde{q}_{sab}
\left(\sum_{i}v_iw_{ia}w_{ib}-Nq_{sab}\right)
\right\}.
\eea
As the order parameters, we define
\bea
q_{wab}\eq\f{1}{N}\sum_{i=1}^Nw_{ia}w_{ib},\\
q_{sab}\eq\f{1}{N}\sum_{i=1}^Nv_iw_{ia}w_{ib},
\eea
and $\tilde{q}_{wab}$ and 
$\tilde{q}_{sab}$ are the corresponding auxiliary parameters.
Moreover, 
$Q_w=\left\{q_{wab}\right\}\in{\bf R}^{n\times n}$, 
$Q_s=\left\{q_{sab}\right\}\in{\bf R}^{n\times n}$,
$\tilde{Q}_w=\left\{\tilde{q}_{wab}\right\}\in{\bf R}^{n\times n}$, and 
$\tilde{Q}_s=\left\{\tilde{q}_{sab}\right\}\in{\bf R}^{n\times n}$
are used. In addition, using the Gaussian integral with respect to $\vec{u}_a,\vec{z}_a$,
\bea
&&
\f{1}{(2\pi)^{pn}}
\area \prod_{a}d\vec{u}_a
d\vec{z}_a
\exp\left(
-\f{\b}{2}\sum_{\mu,a}z_{\mu a}^2
\right.\nn
&&
\left.
+i\sum_{\mu,a}u_{\mu a}z_{\mu a}
-\f{1}{2}\sum_{\mu,a,b}q_{sab}u_{\mu a}u_{\mu b}\right)\nn
\eq
\exp
\left(-\f{p}{2}\log\det\left|I+\b Q_s\right|\right),
\eea
where $I$ is the ${n\times n}$  identify matrix.
In a similar way, using the Gaussian integral 
with respect to  $\vec{w}_a$,
\bea
&&
\f{1}{(2\pi)^{\f{Nn}{2}}}
\area\prod_ad\vec{w}_a
\exp\left(
-\f{1}{2}\sum_{i,a,b}\tilde{q}_{wab}w_{ia}w_{ib}
\right.\nn
&&\left.
-\f{1}{2}\sum_{i,a,b}\tilde{q}_{sab}v_iw_{ia}w_{ib}
+\sum_{i,a}k_aw_{ia}
+\sum_{i,a}\theta_ar_iw_{ia}\right)\nn
\eq
\exp\left(
-\f{1}{2}\sum_{i}\log\det\left|\tilde{Q}_w+v_i\tilde{Q}_s\right|
\right.\nn
&&\left.
+\f{1}{2}
\sum_i(\vec{k}+r_i\vec{\theta})^{\rm T}
\left(\tilde{Q}_w+v_i\tilde{Q}_s\right)^{-1}
(\vec{k}+r_i\vec{\theta})
\right).
\eea
From this, 
\bea
&&\log E
\left[Z^n(R,X,\vec{r})\right]\nn
\eq\mathop{\rm Extr}_{\vec{k},\vec{\theta},Q_w,\tilde{Q}_w,Q_s,\tilde{Q}_s}
\left\{
\f{N}{2}\sum_{a,b}q_{wab}\tilde{q}_{wab}
+\f{N}{2}\sum_{a,b}q_{sab}\tilde{q}_{sab}
\right.
\nn
&&-N\sum_ak_a
-NR\sum_a\theta_a
-\f{p}{2}\log\det\left|I+\b Q_s\right|
\nn
&&
-\f{1}{2}\sum_{i}\log\det\left|\tilde{Q}_w+v_i\tilde{Q}_s\right|
\nn
&&
\left.
+\f{1}{2}
\sum_i(\vec{k}+r_i\vec{\theta})^{\rm T}
\left(\tilde{Q}_w+v_i\tilde{Q}_s\right)^{-1}
(\vec{k}+r_i\vec{\theta})
\right\}.
\eea
and, in the limit of a large number of assets,
using the replica symmetric solution derived in Eqs. (\ref{eq11-1}) to
 (\ref{eq16-1}),
\bea
&&
\lim_{N\to\infty}\f{1}{N}
\log E
\left[Z^n(R,X,\vec{r})\right]\nn
\eq\mathop{\rm Extr}_\Theta
\left\{
\f{n}{2}(\chi_w+q_w)(\tilde{\chi}_w-\tilde{q}_w)
-\f{n(n-1)}{2}q_w\tilde{q}_w
\right.\nn
&&
+\f{n}{2}(\chi_s+q_s)(\tilde{\chi}_s-\tilde{q}_s)
-\f{n(n-1)}{2}q_s\tilde{q}_s
-nk-nR\theta
\nn
&&-\f{\a(n-1)}{2}\log(1+\b\chi_s)
-\f{\a}{2}\log(1+\b\chi_s+n\b q_s)
\nn
&&-\f{n-1}{2}\left\langle
\log(\tilde{\chi}_w+v\tilde{\chi}_s)
\right\rangle\nn
&&
-\f{1}{2}\left\langle
\log(\tilde{\chi}_w+v\tilde{\chi}_s-n(\tilde{q}_w+v\tilde{q}_s)
)
\right\rangle\nn
&&
\left.
+\f{n}{2}
\left\langle
\f{(k+r\theta)^2}
{\tilde{\chi}_w+v\tilde{\chi}_s-n(\tilde{q}_w+v\tilde{q}_s)}
\right\rangle
\right\}
\eea
can be calculated. Substituting the result into \sref{eq9},
\sref{eq17} is obtained by using $\a=p/N\sim O(1)$.

By a similar argument, 
we can easily solve the dual problem in subsection \ref{sec4.2} by using replica analysis.
From the discussion in previous work {\cite{Shinzato-2016-PRE11}}, 
the partition function $Z(\ve,X,\vec{r})$ and 
the Hamiltonian ${\cal H}'(\vec{w}|\vec{r})$
are defined as follows:
\bea
Z(\ve,X,\vec{r})\eq\int_{\vec{w}\in{\cal W}'}
d\vec{w}e^{\b{\cal H}'(\vec{w}|\vec{r})},\\
{\cal H}'(\vec{w}|\vec{r})\eq\sum_{i=1}^Nr_iw_i,
\eea
where the feasible portfolio subset space characterized by 
the constraints of budget and investment risk,
\bea
{\cal W}'\eq\left\{\vec{w}\in{\bf R}^N
\left|
\vec{w}^{\rm T}\vec{e}=N,
N\ve=\f{1}{2}\vec{w}^{\rm T}J\vec{w}
\right.\right\},\qquad
\eea
is employed. From this, using the self-averaging property of this disordered system,
 in order to perform this optimization 
problem, 
\bea
\phi\eq\lim_{N\to\infty}\f{1}{N}E
\left[\log Z(\ve,X,\vec{r})\right],
\eea
is defined. Then, from the following identical equations,
\bea
R^{\max}\eq\lim_{\b\to\infty}\pp{\phi}{\b},\\
R^{\min}\eq\lim_{\b\to-\infty}\pp{\phi}{\b},
\eea
the maximal and minimal expected returns per asset, $R^{\max}$ and 
$R^{\min}$, can be evaluated.

In a similar way to the above-discussed replica analysis,
using the replica symmetric solution,
\bea
\label{eq-a16}
\phi\eq\mathop{\rm Extr}_\Theta
\left\{
\ve\theta-k-\f{\a}{2}\log(1+\theta\chi_s)
-\f{\a\theta q_s}{2(1+\theta\chi_s)}
\right.
\nn &&+\f{1}{2}(\chi_w+q_w)(\tilde{\chi}_w-\tilde{q}_w)+\f{1}{2}q_w\tilde{q}_w
\nn &&+\f{1}{2}(\chi_s+q_s)(\tilde{\chi}_s-\tilde{q}_s)+\f{1}{2}q_s\tilde{q}_s
+\f{1}{2}\left\langle\f{\tilde{q}_w+v\tilde{q}_s}{\tilde{\chi}_w+v\tilde{\chi}_s}\right\rangle
\nn &&
\left.
-\f{1}{2}\left\langle\log(\tilde{\chi}_w+v\tilde{\chi}_s)\right\rangle
+\f{1}{2}\left\langle\f{(k+r\b)^2}{\tilde{\chi}_w+v\tilde{\chi}_s}\right\rangle
\right\},
\eea
can also be {estimated}  where $\Theta=\left\{k,\theta,\chi_w,q_w,\tilde{\chi}_w,\tilde{q}_w,
\chi_s,q_s,\tilde{\chi}_s,\tilde{q}_s
\right\}$ is the set of the order 
parameters. {From the extremum conditions for \sref{eq-a16} with respect to these parameters,}
in terms of parameter $\theta$, 
the primal parameters are as follows:
\bea
\chi_s\eq\f{1}{\theta(\a-1)},\\
q_s\eq\f{\a}{(\a-1)\left\langle v^{-1}\right\rangle}
+\f{\a\left\langle v^{-1}\right\rangle V_1}{(\a-1)^3}\left(\f{\b}{\theta}\right)^2,\\
k\eq\f{\theta(\a-1)}{\left\langle v^{-1}\right\rangle}-\b R_1.
\eea
Furthermore,
\bea
\label{eq-a20}
\pp{\phi}{\b}
\eq\f{\b
\left\langle v^{-1}r^2\right\rangle
}{\theta(\a-1)}+
\f{k\left\langle v^{-1}r\right\rangle}{\theta(\a-1)}\nn
\eq R_1+\f{
\left\langle v^{-1}\right\rangle V_1
}{\a-1}\f{\b}{\theta},
\eea
is obtained. Then in order to 
analyze the upper and lower bounds of the expected return per asset, we need to assess $\theta$, which 
{needs to satisfy}  the following equation:
\bea
\ve\eq
\f{\a\chi_s}{2(1+\theta\chi_s)}
+\f{\a q_s}{2(1+\theta\chi_s)^2}\nn
\eq
\f{1}{2\theta}+
\f{\a-1}{2\left\langle v^{-1}\right\rangle}+
\f{\left\langle v^{-1}\right\rangle V_1}{2(\a-1)}
\left(\f{\b}{\theta}\right)^2.
\eea
Rearranging, this can be written as
\bea
\left(\f{\b}{\theta}\right)^2
\eq\f{2(\a-1)}{\left\langle v^{-1}\right\rangle V_1}
\left(
\ve-\f{1}{2\theta}-\f{\a-1}{2\left\langle v^{-1}\right\rangle}
\right).
\eea
Considering the limit as $\left|\b\right|\to\infty$, we assume 
$\b/\theta\sim O(1)$; then 
\bea
\f{\b}{\theta}
\eq\pm\f{\a-1}{\left\langle v^{-1}\right\rangle}
\sqrt{\f{1}{V_1}
\left(
\f{2\left\langle v^{-1}\right\rangle}{\a-1}
\ve-1
\right)}.
\eea
Note that for $\b\to\infty$, the right-hand side must be positive, whereas if 
$\b\to-\infty$, it must be negative.
Substituting this expression into \sref{eq-a20},
we obtain \bea
\lim_{|\b|\to\infty}
\pp{\phi}{\b}
\eq 
\left\{
\begin{array}{ll}
R_1+
\sqrt{V_1
\left(
\f{2\left\langle v^{-1}\right\rangle}{\a-1}
\ve-1
\right)}&\b\to\infty\\
R_1-
\sqrt{V_1
\left(
\f{2\left\langle v^{-1}\right\rangle}{\a-1}
\ve-1
\right)}&\b\to-\infty
\end{array}
\right..\nn
\eea
Thus, $R^{\max}$ and 
$R^{\min}$ are consistent with 
Eqs. (\ref{eq39-1}) and (\ref{eq40-1}).

\section{Replica analysis for moments\label{app-a}}
Here $\f{1}{N}\vec{e}^{\rm T}J^{-1}\vec{e},
\f{1}{N}\vec{r}^{\rm T}J^{-1}\vec{e},
$ and $
\f{1}{N}\vec{r}^{\rm T}J^{-1}\vec{r}
$ are analyzed. First, 
in order to determine them, the partition function $Z(k,\theta,X)$ is defined as follows:
\bea
Z(k,\theta,X)\eq
\f{1}{(2\pi)^{\f{N}{2}}}
\area d\vec{w}
e^{-\f{1}{2}\vec{w}^{\rm T}J\vec{w}+\vec{w}^{\rm T}(k\vec{e}+\theta\vec{r})},
\qquad
\eea
where $J=XX^{\rm T}\in{\bf R}^{N\times N}$. The partition function is calculated using
\bea
\log Z(k,\theta,X)
\eq\f{1}{2}\log\det|J|
+\f{k^2}{2}\vec{e}^{\rm T}J^{-1}\vec{e}
+\f{\theta^2}{2}\vec{r}^{\rm T}J^{-1}\vec{r}\nn
&&+k\theta\vec{r}^{\rm T}J^{-1}\vec{e}.
\eea
From the self-averaging property, in the limit of large $N$,
\bea
\phi(k,\theta)
\eq\lim_{N\to\infty}\f{1}{N}
E\left[
\log Z(k,\theta,X)
\right],
\eea
from {the second derivatives} of $\phi$ with respect to $k,\theta$,
the typical behaviors of 
$\f{1}{N}\vec{e}^{\rm T}J^{-1}\vec{e},
\f{1}{N}\vec{r}^{\rm T}J^{-1}\vec{e}$, and
$\f{1}{N}\vec{r}^{\rm T}J^{-1}\vec{r}
$ are easily determined. Here, 
using replica analysis and the replica symmetric solution in the limit that  
the number of assets $N$ is large,
\bea
\label{eq-b333}
\phi(k,\theta)
\eq\mathop{\rm Extr}_{\chi_s,q_s,\tilde{\chi}_s,\tilde{q}_s}
\left\{
\f{1}{2}(\chi_s+q_s)(\tilde{\chi}_s-\tilde{q}_s)+\f{q_s\tilde{q}_s}{2}
\right.\nn
&&-\f{\a}{2}\log(1+\chi_s)
-\f{\a q_s}{2(1+\chi_s)}
-\f{1}{2}\left\langle
\log v
\right\rangle\nn
&&
\left.
-\f{1}{2}\log\tilde{\chi}_s+\f{\tilde{q}_s}{2\tilde{\chi}_s}
+\f{1}{2\tilde{\chi}_s}\left\langle
\f{(k+r\theta)^2}{v}
\right\rangle
\right\},\qquad
\eea
is obtained. {From the extremum conditions for \sref{eq-b333}, }
\bea
\chi_s\eq\f{1}{\a-1},\\
q_s\eq\f{\a}{(\a-1)^3}
\left\langle
\f{(k+r\theta)^2}{v}
\right\rangle,\\
\tilde{\chi}_s\eq\a-1,\\
\tilde{q}_s\eq\f{1}{\a-1}
\left\langle
\f{(k+r\theta)^2}{v}
\right\rangle,
\eea
are obtained by using the replica symmetric solution 
in Eqs. (\ref{eq11}) and (\ref{eq13}). {Plugging these into \sref{eq-b333}, }
\bea
\phi(k,\theta)\eq\f{1}{2}
-\f{\a}{2}\log\f{\a}{\a-1}-\f{1}{2}\log(\a-1)\nn
&&
-\f{1}{2}\left\langle
\log v
\right\rangle+
\f{1}{2(\a-1)}\left\langle
\f{(k+r\theta)^2}{v}
\right\rangle,
\eea
is obtained. Thus, 
$\f{1}{N}\vec{e}^{\rm T}J^{-1}\vec{e},
\f{1}{N}\vec{r}^{\rm T}J^{-1}\vec{e},
$ and $
\f{1}{N}\vec{r}^{\rm T}J^{-1}\vec{r}
$ are calculated as follows:
\bea
\label{eq-a10}
\lim_{N\to\infty}\f{1}{N}\vec{e}^{\rm T}J^{-1}
\vec{e}\eq\pp{^2\phi(k,\theta)}{k^2}\nn
\eq\f{\left\langle v^{-1}\right\rangle}{\a-1},\\
\label{eq-a11}
\lim_{N\to\infty}\f{1}{N}\vec{r}^{\rm T}J^{-1}
\vec{e}\eq\pp{^2\phi(k,\theta)}{\theta\p k}\nn
\eq\f{\left\langle v^{-1}r\right\rangle}{\a-1},\\
\label{eq-a12}
\lim_{N\to\infty}\f{1}{N}\vec{r}^{\rm T}J^{-1}
\vec{r}\eq\pp{^2\phi(k,\theta)}{\theta^2}\nn
\eq\f{\left\langle v^{-1}r^2\right\rangle}{\a-1}.
\eea
\section{Stochastic Optimization\label{app-b}}
In this appendix, we summarize the framework of stochastic optimization \cite{Shinzato-2015-PLOS7}.
First, for a given random variable $X$, 
using a real-valued function bounded below with respect to the control parameter
 $w\in{\cal W}$, 
$f(w,X)$, we discuss the optimal solution $w$ which can minimize 
$f(w,X)$, the minimal value of $f(w,X)$, and 
its typical behavior. The random variable $X$ is 
assumed to follow one of the well-known distributions and the feasible subset space of the control parameter $w$ is 
${\cal W}$.
From the {below} discussion, the following results do not always require that 
$f(w,X)$ {is} convex with respect to $w$. 

For a pair $w,X$,
\bea
\label{eq-b1}
f(w,X)\ge \mathop{\min}_{w\in{\cal W}}
f(w,X),
\eea
holds. Let $w^*(X)$ be the value of $w$ which realizes the minimum on the right-hand side, that is,
\bea
\label{eq-b2}
w^*(X)\eq\arg
\mathop{\min}_{w\in{\cal W}}
f(w,X).
\eea
Thus, the equality case of \sref{eq-b1} can be rewritten as follows:
\bea
\label{eq-b3}
f(w^*(X),X)\eq
\mathop{\min}_{w\in{\cal W}}
f(w,X),
\eea
that is, 
\bea
\label{eq-b4}
f(w,X)\ge 
f(w^*(X),X),
\eea
in which one should note that the optimal solution $w^*(X)$ 
depends on random variable $X$, as indicated by the notation.

Next we {can} take the expectation of both sides of \sref{eq-b4} 
with respect to random variable $X$:
\bea
\label{eq-b5}
E_X[f(w,X)]
\ge
E_X[f(w^*(X),X)].
\eea
Since the right-hand side of \sref{eq-b5} is 
constant and the left-hand side holds for any control parameter $w\in{\cal W}$, 
the following inequality holds:
\bea
\mathop{\min}_{w\in{\cal W}}
E_X[f(w,X)]
\ge
E_X\left[f(w^*(X),X)\right].
\eea
We can substitute \sref{eq-b3} into this right-hand side to also obtain
\bea
\label{eq-b7}
\mathop{\min}_{w\in{\cal W}}
E_X[f(w,X)]
\ge
E_X\left[\mathop{\min}_{w\in{\cal W}}f(w,X)\right].
\eea
Thus,
the minimum of the expectation of 
$f(w,X)$ 
with respect to control parameter $w$,
$\mathop{\min}_{w\in{\cal W}}
E_X[f(w,X)]$, is 
not always less than 
the expectation of the minimum of $f(w,X)$ with respect to $w$, 
$E_X\left[\mathop{\min}_{w\in{\cal W}}f(w,X)\right]$.
Further, by a similar argument,
we can also consider the maximization of a real-valued 
{function bounded above with respect to $w$, $g(w,X)$, and  obtain}
\bea
\label{eq-b7-1}
\mathop{\max}_{w\in{\cal W}}
E_X[g(w,X)]
\le
E_X\left[\mathop{\max}_{w\in{\cal W}}g(w,X)\right].
\eea

{Returning to the minimization problem, }
suppose $\mathop{\min}_{w\in{\cal W}}f(w,X)$ satisfies the following 
self-averaging property:
\bea
\label{eq-b8}
\mathop{\min}_{w\in{\cal W}}f(w,X)
\eq
E_X\left[\mathop{\min}_{w\in{\cal W}}f(w,X)\right].
\eea
Then from Eqs. (\ref{eq-b7}) and 
(\ref{eq-b8}),
\bea
\label{eq-b9}
\mathop{\min}_{w\in{\cal W}}
E_X[f(w,X)]
\ge
\mathop{\min}_{w\in{\cal W}}f(w,X),
\eea
{which was} discussed in Ref.~\cite{Shinzato-2015-PLOS7}. That is, 
in terms of the discussion in the main text, 
control parameter $w$ is a portfolio, 
random variable $X$ is a return rate matrix,
real-valued function $f(w,X)$ bounded from below is 
the investment risk, and 
the feasible subset space ${\cal W}$ corresponds to 
several constraints on the portfolio. 
Thus, the discussion here 
clarifies that 
the ordinary portfolio which can minimize the expected investment risk discussed in operations research,
$w^{\rm OR}=\arg\mathop{\min}_{w\in{\cal W}}
E_X[f(w,X)]$, is 
not always consistent with 
the optimal portfolio which can minimize the investment risk, 
$w^*(X)=\arg\mathop{\min}_{w\in{\cal W}}f(w,X)$, 
and which is {sought}  by rational investors.
Namely, as a physical interpretation of 
\sref{eq-b9}, 
the left-hand side of \sref{eq-b9} corresponds to an annealed disordered system 
and the right-hand side of \sref{eq-b9} is related to a quenched disordered system.
Moreover, in previous work \cite{Shinzato-2015-PLOS7},
it was verified that the minimal investment risk per asset $\ve$ and its 
investment concentration 
$q_w$ (and Sharpe ratio $S$, which is defined using the minimal investment risk per asset $\ve$
and the expected return coefficient $R$) 
satisfy the self-averaging property.

\end{document}